\documentclass[aps,prl,preprint,groupedaddress]{revtex4-1}

\usepackage{amsmath,amssymb,natbib,bm,graphicx,url}
\usepackage[british]{babel}

\begin{document}


\title{Metamagnetism of weakly-coupled antiferromagnetic topological insulators}



\author{Aoyu Tan$^{1,2}$, Valentin Labracherie$^{1,2}$, Narayan Kunchur$^2$, Anja U.B. Wolter$^2$, Joaquin Cornejo$^2$, Joseph Dufouleur$^{2,3}$, Bernd B\"uchner$^{2,4}$, Anna Isaeva$^{2,4}$, and Romain Giraud$^{1,2}$}
\affiliation{
$^{1}$Univ. Grenoble Alpes, CNRS, CEA, Spintec, F-38000 Grenoble, France\\
$^{2}$Institute for Solid State Physics, Leibniz IFW Dresden, D-01069 Dresden, Germany\\
$^{3}$Center for Transport and Devices, TU Dresden, D-01069 Dresden, Germany\\
$^{4}$Faculty of Physics, TU Dresden, D-01062 Dresden, Germany\\}



\date{\today}

\begin{abstract}

The magnetic properties of the van der Waals magnetic topological insulators MnBi$_2$Te$_4$ and MnBi$_4$Te$_7$ are investigated by magneto-transport measurements. We evidence that the relative strength of the inter-layer exchange coupling $J$ to the uniaxial anisotropy $K$ controls a transition from an A-type antiferromagnetic order to a ferromagnetic-like metamagnetic state. A bi-layer Stoner-Wohlfarth model allows us to describe this evolution, as well as the typical angular dependence of specific signatures, such as the spin-flop transition of the uniaxial antiferromagnet and the switching field of the metamagnet.
\end{abstract}


\maketitle





The coexistence of large spin-orbit and exchange couplings in 3D crystals can lead to a variety of topological electronic phases, some of which being tunable by changing the magnetic order parameter (orientation, amplitude) or the micro-magnetic structure \cite{Li2010,Mong2010,Yasuda2017,Smejkal2018}. 
This requires the accurate control of the magnetic properties however, also at the microscopic level. 
A breakthrough was the discovery of the quantum anomalous Hall (QAH) state in diluted magnetic topological insulators \cite{Chang2013}, with dissipationless edge states induced by the magnetization. Due to a small energy gap of the surface-state band structure, the Hall resistance quantization is only observed at sub-kelvin temperatures \cite{Bestwick2015,Chang2015,Goetz2018,Fox2018}.

Recently, stoichiometric magnets have raised specific interest \cite{Otrokov2017a,Otrokov2017b,Li2019,Tokura2019}, with the possibility to tailor multi-layers of exchange-coupled 2D ferromagnets having a non-trivial band structure and larger gaps. In particular, MnBi$_2$Te$_4$ was evidenced as the first antiferromagnetic topological insulator, with a N\'eel temperature $T_\textrm{N}$=24~K \cite{Li2019,Otrokov2019a,Gong2019,Lee2019,Yan2019a,Chen2019}. 
Besides, novel topological phases and transitions were predicted in antiferromagnets \cite{Li2010,Mong2010,Baireuther2014}, as well as parity effects in thin magnetic multilayers \cite{Otrokov2019,Sun2019}. 
Theoretical predictions also considered other topological phases in the bulk, such as magnetic Weyl semimetals or axion electrodynamics \cite{Otrokov2019,Zhang2019,Li2019a,Wang2019b}. In all cases, the control of a topological state is directly related to that of the micro-magnetic structure, and the quantized Hall state was observed in large magnetic fields only \cite{Deng2019,Liu2019,Ge2019}. 
Importantly, van der Waals multi-layers of 2D ferromagnets offer the possibility to modify the inter-layer exchange coupling $J$, with non-magnetic spacers, whereas the single-layer magnetic anisotropy $K$ remains barely affected. This can be achieved in the so-called MBT family, [MnBi$_2$Te$_4$][Bi$_2$Te$_3$]$_n$ with the integer $n \geqslant 0$, that ideally realizes stoichiometric magnetic topological insulators \cite{Hu2019,Sun2019,Vidal2019a,Yan2019b,Shi2019,Wu2019,Xu2019,Klimovskikh2019,Otrokov2019a}. The magnetic base unit, a single MnBi$_2$Te$_4$ septuple layer, is a 2D ferromagnet (intra-layer coupling $J_\textrm{F}<0$) with a perpendicular anisotropy $K_\textrm{U}$ that stabilizes an out-of-plane ferromagnetic order and generates the QAH state. Stacks of septuple layers form the MnBi$_2$Te$_4$ compound, with an antiferromagnetic inter-layer coupling ($J=J_\textrm{AF}>0$). It is also possible to grow related crystals that have $n$ units of the non-magnetic Bi$_2$Te$_3$ spacer in between 2D ferromagnetic layers, and therefore a reduced exchange coupling $J$. 

In this work, we evidence that such crystalline MBT magnetic multilayers are actually text-book systems that realize the weak-coupling regime of uniaxial anti-ferromagnets, for all compounds but the MnBi$_2$Te$_4$, with robust meta-magnetic properties controlled by their perpendicular anisotropy. To evidence this behavior, we investigated the magnetic properties of Hall-bar shaped nanostructures of both MnBi$_2$Te$_4$ and MnBi$_4$Te$_7$, in a comparative study, by magneto-transport measurements. Below their N\'eel temperature, the typical signature of an A-type collinear antiferromagnet, a 
spin-flop transition, is observed. 
However, MnBi$_4$Te$_7$ undergoes another transition to a bi-stable metamagnetic state at lower temperatures, with a fully-saturated remnant magnetization below about 5K and abrupt spin-flip transitions. This evolution is well described by a magnetic bi-layer Stoner-Wohlfarth model with an inter-layer exchange coupling $J$ and an effective anisotropy $K$ related to the single-layer uniaxial anisotropy $K_\textrm{U}$. Our model also reproduces the angular dependence of these different magnetic states under a tilted magnetic field. This finding of metamagnetism is very general for van der Waals 2D-layered ferromagnets with a weak inter-layer exchange coupling as compared to their uniaxial anisotropy strength. In the limit of a large $K/J$ ratio, the model suggests a direct phase transition from paramagnetism to metamagnetism, with a saturated magnetization at remanence up to the blocking temperature $T_\textrm{B}$ of the 2D ferromagnet base unit, with an upper bond given by the critical temperature of the magnetic base unit of about 11(1)K. 
This situation is probably already realized for $n=2$, that is, for MnBi$_6$Te$_{10}$, which indeed agrees with recent experimental findings, yet interpreted in terms of a ferromagnetic state \cite{Wu2019,Shi2019,Yan2019b}. Our study actually shows the importance of both the \emph{intra-layer} 2D exchange coupling $J_\textrm{F}$ and the \emph{perpendicular anisotropy} $K_\textrm{U}$ to realize robust metamagnetic states and to stabilize the QAH regime at higher temperatures.

Nanoflakes of both MnBi$_2$Te$_4$ and MnBi$_4$Te$_7$ were obtained by mechanical exfoliation of large single crystals, the crystal growth and bulk properties of which were reported elsewhere \cite{Zeugner2019,Souchay2019,Otrokov2019a,Vidal2019a}. Nanostructures were transferred onto a SiO$_2$/Si$^{++}$ substrate, and then further processed by e-beam lithography to prepare Cr/Au ohmic contacts and then shaped into a Hall-bar geometry (using a negative e-beam resist and Ar-ion milling). Magneto-transport measurements were performed with ac lock-in amplifiers, using a small polarization current, down to very low temperatures ($T>$70~mK) and in an Oxford Instr. 3D-vector 2T magnet. High-field measurements, up to 14T, were realized in a variable-temperature insert, down to 1.8K, at different magnetic field orientations by using a mechanical rotator.


Both MnBi$_2$Te$_4$ and MnBi$_4$Te$_7$ nanoflakes showed a dirty metal-like behavior due to disorder (see SI). Moreover, the average carrier mobility is enhanced by spin-dependent scattering at a phase transition to a N\'eel antiferromagnetic state, giving a resistivity peak at the critical temperature $T_\textrm{N}$ (maximum of magnetic fluctuations), with $T_\textrm{N}$=23.5(5)K and $T_\textrm{N}$=12.5(5)K, respectively. A simple mean-field model analysis already reveals the much reduced inter-layer exchange coupling $J_\textrm{AF}$ in MnBi$_4$Te$_7$ compared to MnBi$_2$Te$_4$. The magnetic susceptibility above $T_\textrm{N}$ gives a paramagnetic Weiss temperature $\theta_P=$1(1)K for MnBi$_2$Te$_4$ and $\theta_P=$12(1)K for MnBi$_4$Te$_7$ \cite{Otrokov2019a,Souchay2019,Vidal2019a,Wu2019,Yan2019b}. Since the ratio $\theta_P/T_N$ is given by $[(J_\textrm{F}+J_\textrm{AF})/ (J_\textrm{F}-J_\textrm{AF})]$ \cite{Coey2010}, this shows that $J_\textrm{AF}/J_\textrm{F} \approx -0.92$ for MnBi$_2$Te$_4$ and $J_\textrm{AF}/J_\textrm{F} \approx -0.04$ for MnBi$_4$Te$_7$. 
All MBT-$n$ compounds are therefore weakly-coupled 2D magnetic multilayers ($J_\textrm{AF} \ll \vert J_\textrm{F} \vert$), apart from MnBi$_2$Te$_4$ that has $J_\textrm{AF} \lesssim \vert J_\textrm{F} \vert$. Below $T_\textrm{N}$, the resistivity is reduced upon cooling the sample, as magnons are progressively frozen. Another evidence of the weaker inter-layer coupling in MnBi$_4$Te$_7$ is thus given by the faster decrease of the resistivity with decreasing the temperature, since 2D magnetic fluctuations are efficiently gapped by the uniaxial anisotropy. 

\begin{figure}[!h]
\includegraphics[width=\columnwidth]{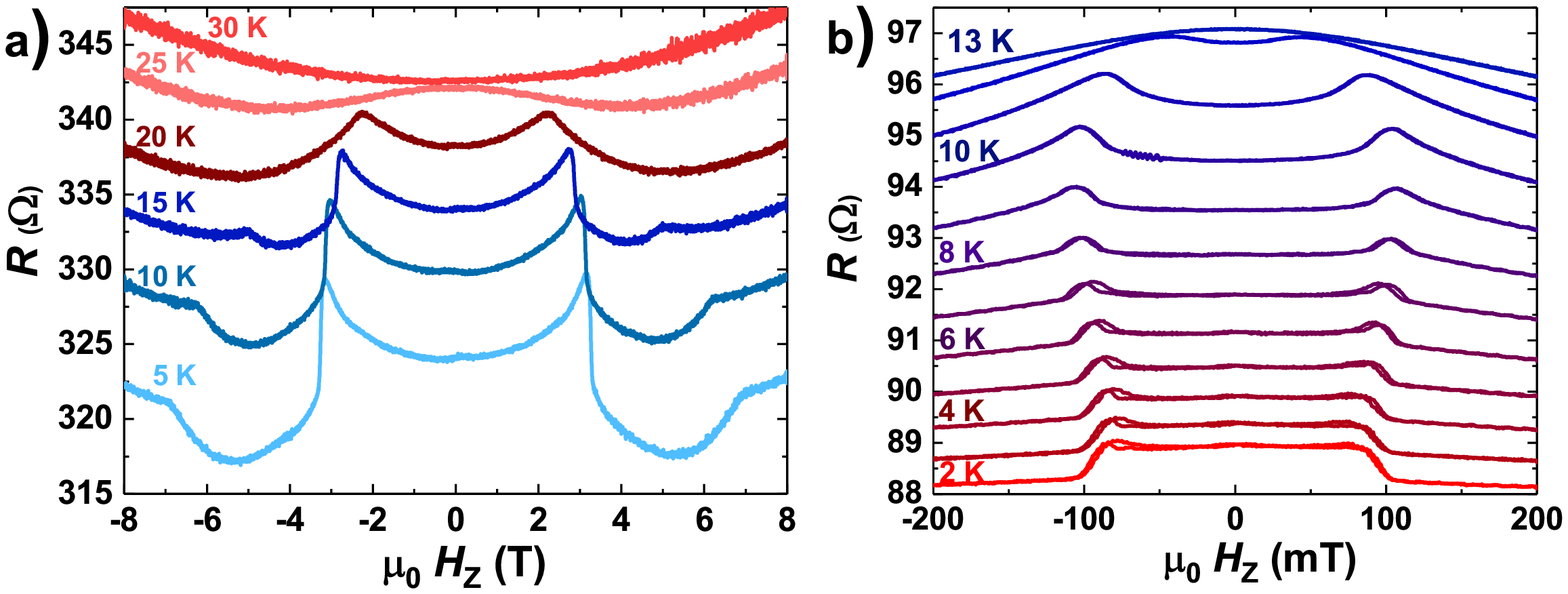}
\centering
\caption{ 
Temperature dependence of the longitudinal magneto-resistance for MnBi$_2$Te$_4$ (\textbf{a}) and MnBi$_4$Te$_7$ (\textbf{b}), with peaks at the magnetization reversal (spin-flop or spin-flip transitions) below $T_\textrm{N}$=23.5(5)K (\textbf{a}) and $T_\textrm{N}$=12.5(5)K (\textbf{b}). 
Curves are shifted for clarity.
}
\label{fig1}
\end{figure}
\begin{figure}[!h]
\includegraphics[width=\columnwidth]{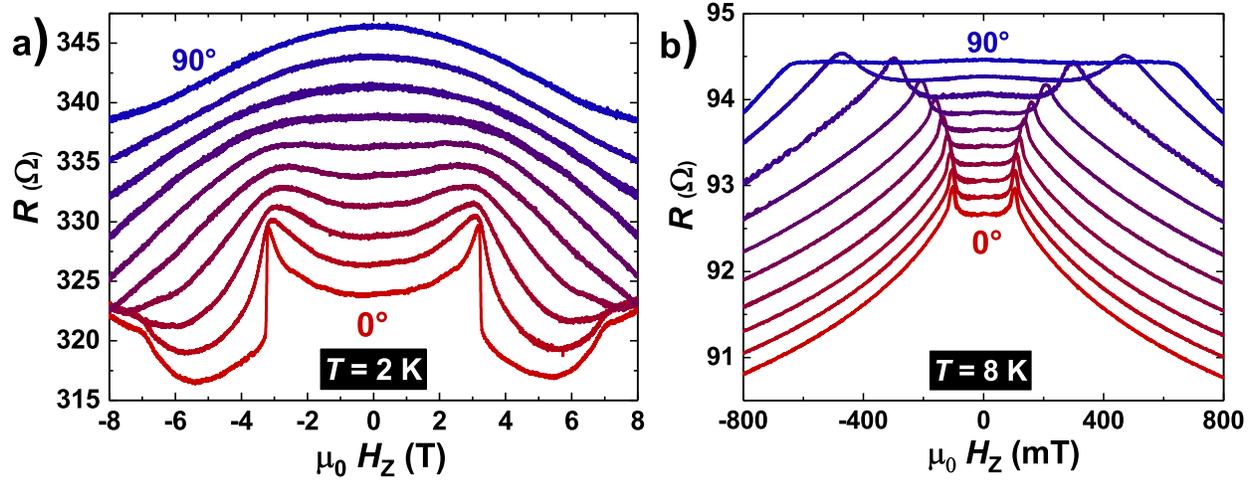}
\centering
\caption{Angular dependence of the magneto-resistance for MnBi$_2$Te$_4$ (\textbf{a}), showing the fast vanishing of the spin-flop transition for all temperatures down to $T=2$~K, and for MnBi$_4$Te$_7$ (\textbf{b}), showing the smooth evolution of the spin-flop transition measured at $T=8$~K. 
Curves with a 10$^{\circ}$ step are shifted for clarity.
}
\label{fig2}
\end{figure}

A clear signature of the collinear A-type antiferromagnetic state is further observed in the longitudinal magneto-resistance. In zero magnetic field, the two sub-lattice magnetizations are aligned along the uniaxial anisotropy axis perpendicular to the septuple plane. If the field is applied along the easy axis, the magneto-resistance shows a reversible curve with a peak at the spin-flop transition, specific to a uniaxial antiferromagnet with a dominant exchange energy, when sub-lattice magnetizations suddenly evolve to a canted state due to the finite antiferromagnetic coupling and anisotropy \cite{Coey2010}. The temperature dependence of the spin-flop field $H_\textrm{SF} \propto \sqrt{JK}$ is related to that of the effective anisotropy $K \sin^2\theta = K_\textrm{U}/M_\textrm{S}^2 * <M_\textrm{Z}^2>$, where $M_\textrm{Z}$ and $M_\textrm{S}/2$ are the sub-layer perpendicular and saturated magnetization, respectively, and $<>$ is the thermal average. As magnetic fluctuations are reduced at lower temperatures, the effective uniaxial anisotropy $K$ increases, and so does the spin-flop field below $T_\textrm{N}$.
 However, there are some striking differences between both magnets at lower temperatures, due to the relative strength of the exchange field $H_\textrm{exch}\propto J$ compared to the anisotropy field $H_\textrm{A}\propto K$. For MnBi$_2$Te$_4$, $J$ is always larger than $K$. This leads to the features seen in Fig.~\ref{fig1}a). 
First, the spin-flop field is smaller than the saturation field, the latter being solely determined by the exchange field if the field is applied along the anisotropy axis. Second, the spin-flop transition induces a large canting of the magnetization with respect to the uniaxial anisotropy direction, which results in a visible contribution from the negative anisotropic magneto-resistance. This evolution of the magnetization is indeed confirmed by that of the anomalous Hall resistance, which is a measure of the magnetization component $M_\textrm{Z}$ perpendicular to the sample plane (see SI). At higher fields, the magnetization slowly realigns towards the anisotropy axis and the resistance increases again, up to the magnetization saturation field that is clearly observed as a kink in the magneto-resistance. At even larger fields, only the quadratic cyclotron magneto-resistance remains. The angular dependence with a tilted magnetic field confirms this scenario, with the rapid vanishing of the spin-flop event and the sole contribution of the anisotropic magneto-resistance at large angles (Fig.~\ref{fig2}a). For MnBi$_4$Te$_7$, we found two different regimes. Below $T_\textrm{N}$ and above about $T=8$~K, the magnetic properties are also those of a uniaxial antiferromagnet. The spin-flop transition is however observed at a much smaller field, as expected due to the reduced inter-layer coupling $J$. Considering the anisotropy is determined locally within a septuple layer, and therefore similar for both compounds, this would give a ratio $H_\textrm{SF}^{124}/ H_\textrm{SF}^{147} \approx 5$, whereas it is about 30 since $B_\textrm{SF}^{124}\approx3$~T and $B_\textrm{SF}^{147}\approx100$~mT. This difference is already a sign that the nature of the magnetization reversal changes in MnBi$_4$Te$_7$, as the anisotropy energy becomes larger than the exchange energy. As a consequence, $H_\textrm{SF}^{147}$ has a smooth angular dependence (Fig.~\ref{fig2}b), and the magneto-resistance peak at large angles is related to the sudden change of $M_\textrm{Z}$ (see SI, Fig.~\ref{fig2SI}b) when the anisotropy energy barrier vanishes. 

\begin{figure}[!h]
\includegraphics[width=\columnwidth]{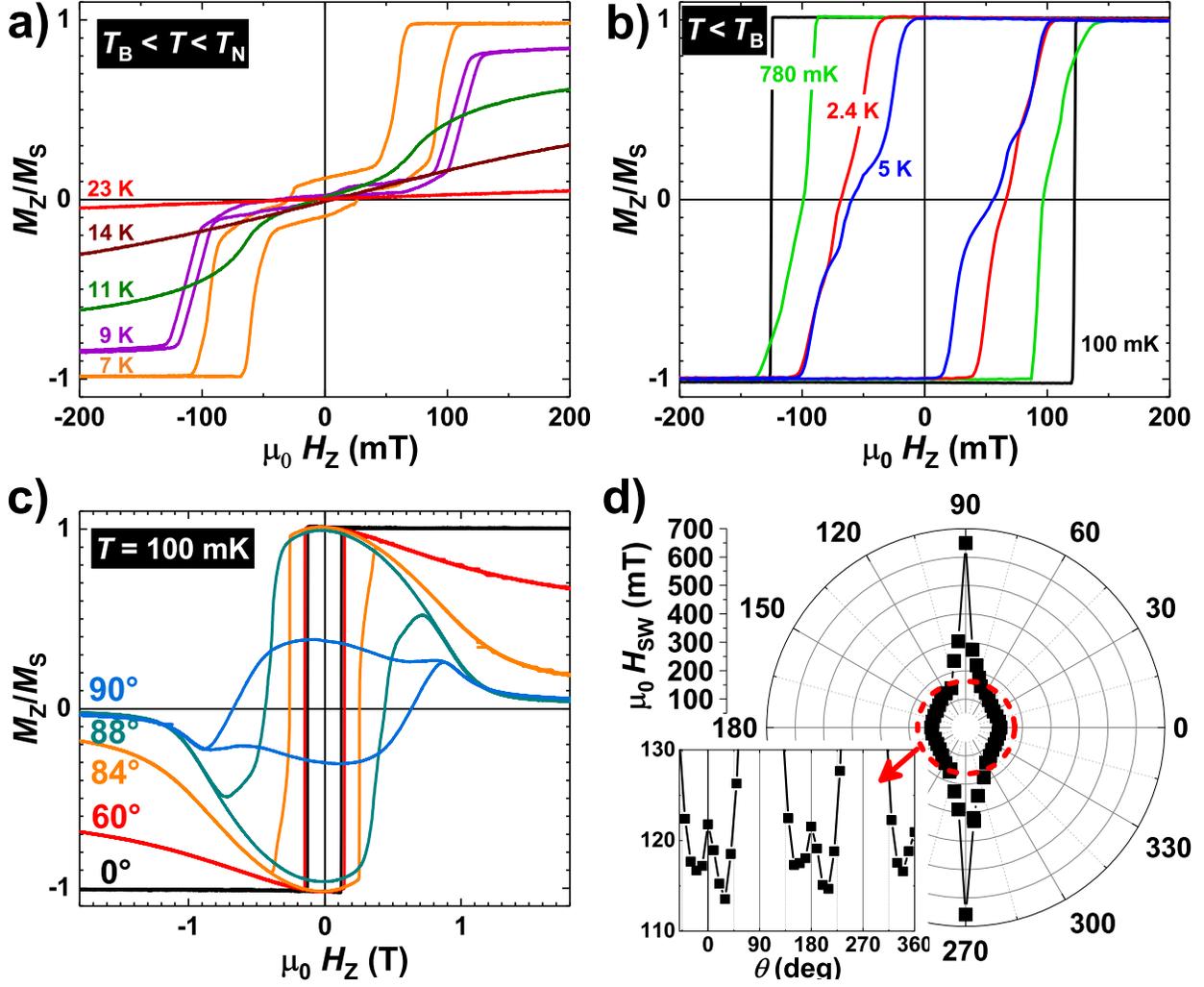}
\centering
\caption{
Perpendicular magnetization $M_\textrm{Z}$ hysteresis loops for MnBi$_4$Te$_7$, normalized to its saturation value $M_\textrm{S}$, showing the split spin-flop transitions in the regime $K<2J$ (\textbf{a}) and the evolution to a meta-magnetic state ($K>2J$) below the blocking temperature $T_\textrm{B}\approx5$~K (\textbf{b}). Another narrower Hall bar shows a perfect spin-flip transition with a well-defined switching field, as shown at $T=100$~mK. The angular dependence of hysteresis loops reveals the dominant influence of the uniaxial anisotropy, with an anisotropy field $\mu_0 H_\textrm{A}\approx 0.7$~T (\textbf{c}). The Stoner-Wohlfarth mechanism is confirmed by the polar plot of the switching field $H_\textrm{SW}$ (\textbf{d}) showing a truncated astroid behavior, with a maximum still clearly seen even along the easy axis (inset).
}
\label{fig3}
\end{figure}

Most important, MnBi$_4$Te$_7$ undergoes a progressive transition at lower temperatures to a metamagnetic phase controlled by the uniaxial anisotropy. We evidence that this evolution of the total out-of-plane magnetization is related to that of the $K/J$ ratio. Contrary to most uniaxial antiferromagnets, for which the exchange energy is much larger than the anisotropy, van der Waals-coupled magnetic multilayers can have competing energies, which results in specific magnetic properties. First, the spin-flop transition becomes hysteretic and two switching fields can be distinguished (Fig.~\ref{fig3}a), as also observed by others \cite{Wu2019}. As shown below by our model, this is due to the relative alignement of the sub-lattice magnetizations, which can be either parallel (P) or antiparallel (AP), resulting in two spin-flop fields $H_\textrm{SF}^\textrm{AP}$ and $H_\textrm{SF}^\textrm{P}$.  
At 5K, the lower switching field $H_\textrm{SF}^\textrm{P}$ changes its sign, and the remnant state becomes fully magnetized (Fig.~\ref{fig3}b). At very low temperatures, the hysteresis loop becomes very sharp with a single switching field (Fig.~\ref{fig3}c), a behavior similar to that of a uniaxial ferromagnet. It is however the spin-flip transition of a metamagnetic state with a dominant uniaxial anisotropy energy ($K \gg J$), as confirmed by the angular dependence of $M_\textrm{Z}$ (Fig.~\ref{fig3}c). Under a tilted field, the saturated magnetization aligns along the applied field, under single-domain rotation, when it compares to the anisotropy field $\mu_0 H_\textrm{A}\approx 0.7$~T. The remnant magnetization remains fully saturated for nearly all angles, but for a large-enough in-plane field that indeed cancels the energy barrier (which thus favors the decomposition in antiferromagnetic domains). Upon field reversal, the switching field $H_\textrm{SW}$ is well defined and, 
after an initial decrease, it has a progressive angular dependence to a maximum value. This upper limit is due to the reduction of the anisotropy energy barrier under a transverse magnetic field. Indeed, the polar plot of $H_\textrm{SW}$ shows the typical profile of a Stoner-Wohlfarth astroid(Fig.~\ref{fig3}d), although it is truncated for small angles, when domain walls can be nucleated by a large-enough $H_\textrm{Z}$ component and the demagnetizing field.

All these experimental results can indeed be explained by a simple model based on two 2D ferromagnetic layers with a uniaxial perpendicular anisotropy $K$ and adding a weak antiferromagnetic exchange coupling $J$, with competing interactions ($K\sim J$). This bi-layer Stoner-Wohlfarth model allows us to describe the evolution from an A-type antiferromagnet to a uniaxial meta-magnet, and it captures the temperature and angular dependences of the magnetization curves as well, given that $K$ decreases with temperature. It also gives values of the $K/J$ ratio required to stabilize each regime. 
We consider the free energy of two magnetic sublattices, each with a uniform magnetization $\overrightarrow{M}_{1,2}=M\overrightarrow{m}_{1,2}$, where $M=M_\textrm{S}/2$ and $\overrightarrow{m}_{1,2}$ are unit vectors.  
For a tilted magnetic field $\overrightarrow{H}$, with a polar angle $\theta$ with respect to the easy-anisotropy axis, each magnetic sublattice has an equilibrium state that can be obtained by minimizing the free energy, where two values $\theta_1$ and $\theta_2$ determine the sub-lattice magnetization orientations. The free energy reads $E = -\mu H[cos(\theta_1-\theta)+ cos(\theta_2-\theta)] + K(sin^2\theta_1 + sin^2\theta_2) + 2J cos(\theta_1-\theta_2)$, where $\mu=\mu_0 M$. 

Using the free energy, we can determine the magnetic ground state for each sub-layer, as well as the energy barrier separating the parallel and antiparallel configurations (see SI). Neglecting thermal fluctuations (which contribute to a finite but small in-plane magnetic susceptibility), the total magnetization is thus calculated for any orientation and amplitude of the applied field.


To evidence the relative influence of the uniaxial anisotropy and of the antiferromagnetic inter-layer exchange coupling, we consider the three limiting cases of a dominant exchange coupling ($K \ll J$), competing couplings ($K \approx 2J$) and a dominant uniaxial anisotropy ($K \gg J$). The magnetization curves along the anisotropy axis $M_z (H_z)$ are thus shown for three $K/J$ ratios, representative of the different regimes  .

\begin{figure}[!h]
\includegraphics[width=\columnwidth]{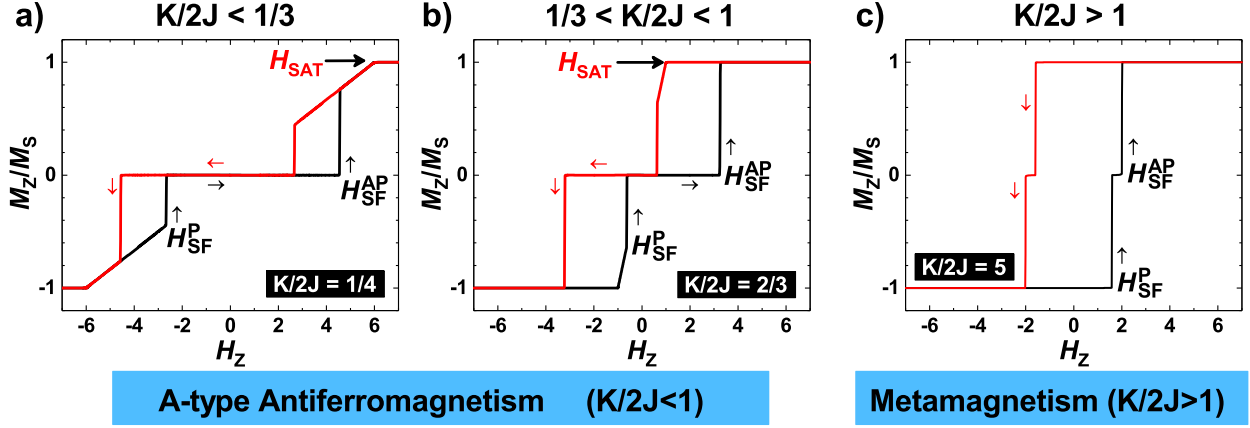}
\centering
\caption{
Calculated hysteresis loops $M_\textrm{Z}(H_\textrm{Z})$ for three $K/2J$ ratios ($\frac{1}{4},\frac{2}{3},5$), representative of the different regimes. An A-type antiferromagnet with competing interactions has two spin-flop transition fields $H_\textrm{SF}^\textrm{AP}$ and $H_\textrm{SF}^\textrm{P}$, depending on the relative alignement of the sub-lattice magnetizations. These can be smaller than the saturation field $H_\textrm{SAT}$ (\textbf{a}, dominant exchange energy) or give a larger hysteresis if the $K/J$ ratio increases, with $H_\textrm{SAT}< H_\textrm{SF}^\textrm{AP}$ and reduced canting (\textbf{b}). The fully-saturated meta-magnetic state develops when the anisotropy energy becomes dominant (\textbf{c}), so that $H_\textrm{SF}^\textrm{P}$ becomes negative and the magnetization switching proceeds as two spin-flip transitions (no canting) in a narrow field range. For a large-enough $K/J$ ratio, the magnetization reversal proceeds as a single spin-flip transition, only controlled by the uniaxial anisotropy.
}  
\label{fig4}
\end{figure}

We first focus on the regime $K/2J<1$, for which spin-flop fields have the same sign. 
The ground state is that of a uniaxial antiferromagnet, with a zero net magnetization. 
By appling a magnetic field along the easy axis, there is a transition at $H_\textrm{SF}^\textrm{AP}$. Depending on the $K/2J$ ratio, the new equilibrium changes either to canted sub-layer magnetization states or to a ferromagnetic-like alignement in a finite field. 
For $K/2J < 1/3$ (Fig.~\ref{fig4}a), the antiferromagnetic ground state undergoes a spin-flop transition to a canted state. Increasing the field progressively brings the staggered magnetizations back to a parallel state, by coherent rotation (linear variation of the $M_z$ component), with a full alignment at the saturation field $H_\textrm{SAT} \approx H_\textrm{exch}$. 
Due to the 
energy barrier, the spin-flop field depends on the relative orientation of the sublattice magnetizations (parallel P or anti-parallel AP), which gives two different fields $H_\textrm{SF}^\textrm{AP}$ and $H_\textrm{SF}^\textrm{P}$. 
For $1/3<K/2J<1$ (Fig.~\ref{fig4}b), the antiferromagnetic ground state undergoes a spin-flip transition to a fully-aligned state. This happens when $H_\textrm{SAT}$ becomes smaller than $H_\textrm{SF}^\textrm{AP}$ (decrease of $J$ and/or increase of $K$).

Upon increasing the $K/J$ ratio, the lower field $H_\textrm{SF}^\textrm{P}$ is progressively reduced, as found in the intermediate regime of MnBi$_4$Te$_7$ ($K$ increases at lower temperatures). For $K/2J=1$, $H_\textrm{SF}^\textrm{P}$ changes its sign, so that the remnant magnetization remains fully magnetized after an initial saturation. Due to the temperature dependence of $K$, this allows us to define a blocking temperature $T_\textrm{B}$ as $H_\textrm{SF}^\textrm{P}(T_\textrm{B})=0$, the condition for a saturated remnant magnetization.  
By further increasing the $K/J$ ratio (Fig.~\ref{fig4}c), a larger hysteresis loop develops, as $H_\textrm{SF}^\textrm{P}$ changed its sign and both spin-flop fields increase ($\vert H_\textrm{SF}^\textrm{P}\vert$ progressively increases faster, up to $H_\textrm{SF}^\textrm{AP}$ in the $K \gg J$ limit). 
This is shown for $K/2J=5$, where the magnetization reversal now proceeds as a narrow double step, which is then the spin-flip transition of a meta-magnet. 
The limit $K/J \gg 1$ is the standard Stoner-Wohlfarth model with a single-step magnetization reversal, for which the switching field is solely controlled by the anisotropy barrier.

This evolution is typical for the magnetic behavior found in MnBi$_4$Te$_7$ (Fig.~\ref{fig3}a,b). At very-low temperature, the magnetization reversal is a direct spin-flip transition that is mostly controlled by the anisotropy. This is confirmed by the angular dependence of the switching field that shows an astroid-like behavior (Fig.~\ref{fig3}d), typical of magnetic systems with a uniaxial anisotropy. The asteroid is well reproduced in the hard-axis direction (evolution of the anisotropy energy barrier with an in-plane applied field), whereas it is truncated in the easy-axis direction, probably due to the formation of domain walls during the magnetization reversal in micron-sized magnets. 
Despite some intrinsic limitations of the single-domain approach \cite{Dieny1990a}, this bi-layer model is very predictive since large domain sizes can be obtained in antiferromagnets, so that the free-energy description captures the physics of the competition between the uniaxial anisotropy and the inter-layer antiferromagnetic exchange coupling.   


In a comprehensive study of the magnetization reversal processes of magnetic topological insulators MnBi$_2$Te$_4$ ($n$=0) and MnBi$_4$Te$_7$ ($n$=1), we evidenced the anisotropy-controlled transition from an A-type collinear antiferromagnet to a fully-saturated meta-magnetic state in weakly-coupled magnetic multi-layers. Based on a simple Stoner-Wohlfarth model modified for a bi-layer system with an antiferromagnetic exchange energy $J$, we reveal that ferromagnetic-like hysteresis loops are actually the signature of a dominant anisotropy energy $K$, which offers the possibility to stabilize a uniform magnetization. For a vanishing inter-layer coupling, as already achieved for MnBi$_6$Te$_{10}$ ($n$=2), the magnetization is that of independent anisotropic 2D ferromagnets.    
Importantly, the detailed understanding of the different ground states of layered magnetic topological insulators is necessary so as to control novel topological states, induced by exchange fields, that can still be tunable by small external fields.

\begin{acknowledgments}

We acknowledge the funding of the European Commission via the TOCHA project H2020-FETPROACT-01-2018 under Grant Agreement 824140.
This work was supported by the German Research Foundation (DFG) in the framework of the SPP 1666 "Topological Insulators" program, of the CRC "Correlated Magnetism - From Frustration to Topology" (SFB-1143, Project No. 247310070), and of the W\"urzburg-Dresden Cluster of Excellence on Complexity and Topology in Quantum Matter - \emph{ct.qmat} (EXC 2147, Project No. 39085490).

\end{acknowledgments}



\vspace{-5mm}

\newpage

\newpage

\newpage

\newpage

\newpage

\section{Supplementary Information: Experiments}

\subsection{Temperature dependence of the longitudinal resistance of MnBi$_2$Te$_4$ and MnBi$_4$Te$_7$ Hall bars}

\begin{figure}[!t!h]
\includegraphics[width=\columnwidth]{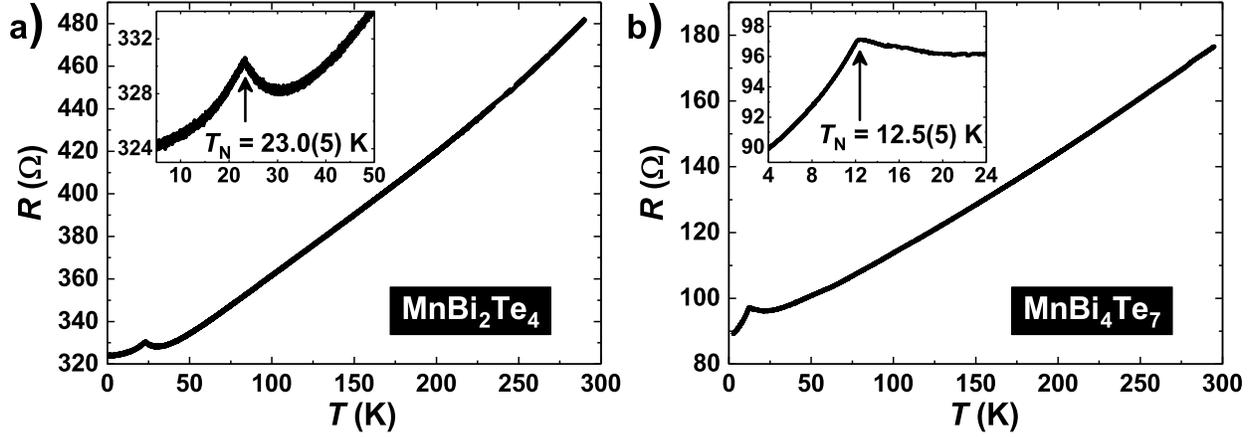}
\centering
\caption{
\textbf{a}), Temperature dependence of the longitudinal resistance of a MnBi$_2$Te$_4$ Hall bar patterned from an exfoliated flake. 
Inset: Enhanced magnetic fluctuations at the N\'eel transition $T_\textrm{N}$=23.5(5)K give a resistance peak, followed by a decrease due to the freezing of magnons; 
\textbf{b}), Temperature dependence of the longitudinal resistance of a MnBi$_4$Te$_7$ Hall bar patterned from an exfoliated flake. 
Insets: Enhanced magnetic fluctuations at the N\'eel transition $T_\textrm{N}$=12.5(5)K give a resistance peak, followed by a decrease due to the freezing of magnons.
}
\label{fig1SI}
\end{figure}

Hall-bar patterned nanostructures were prepared by mechanical exfoliation of MnBi$_2$Te$_4$ and MnBi$_4$Te$_7$ single crystals and e-beam lithography. Magneto-transport measurements were performed in a four-probe configuration, with a well-defined geometry. Both MnBi$_2$Te$_4$ and MnBi$_4$Te$_7$ nanoflakes show a dirty metal-like behavior due to disorder, as evidenced by the temperature dependence of the longitudinal resistance (Fig.~\ref{fig1SI}). In particular, the small value of the residual resistance ratio $RRR=R(300$~K$)/R(4$~K) is typical for the parent compound Bi$_2$Te$_3$, due to scattering related to point defects.


In these magnetic compounds, the mobility is even further reduced by a small degree of inter-mixing between nearby Mn and Bi planes. This results in a smaller value $RRR\approx1.5$ for MnBi$_2$Te$_4$ than for MnBi$_4$Te$_7$ that has a value $RRR\approx2$ closer to that of Bi$_2$Te$_3$. Such a difference is expected since MnBi$_4$Te$_7$ has non-magnetic Bi$_2$Te$_3$ spacers that reduce the relative influence of Mn/Bi inter-mixing on the average mobility, whereas MnBi$_2$Te$_4$ has none. \\

\subsection{Angular dependence of the transverse magneto-resistance \\of MnBi$_2$Te$_4$ and MnBi$_4$Te$_7$ Hall bars}

For both MnBi$_2$Te$_4$ and MnBi$_4$Te$_7$, the transverse resistance has a linear field dependence that corresponds to the normal Hall effect, with a maximum or a vanishing slope if the magnetic field is applied perpendicular to or within the sample plane, respectively. Due to the anomalous Hall effect, it also reveals another contribution related directly to the perpendicular component of the magnetization $M_\textrm{Z} \propto V^\textrm{AHE}$.

\begin{figure}[!h]
\includegraphics[width=\columnwidth]{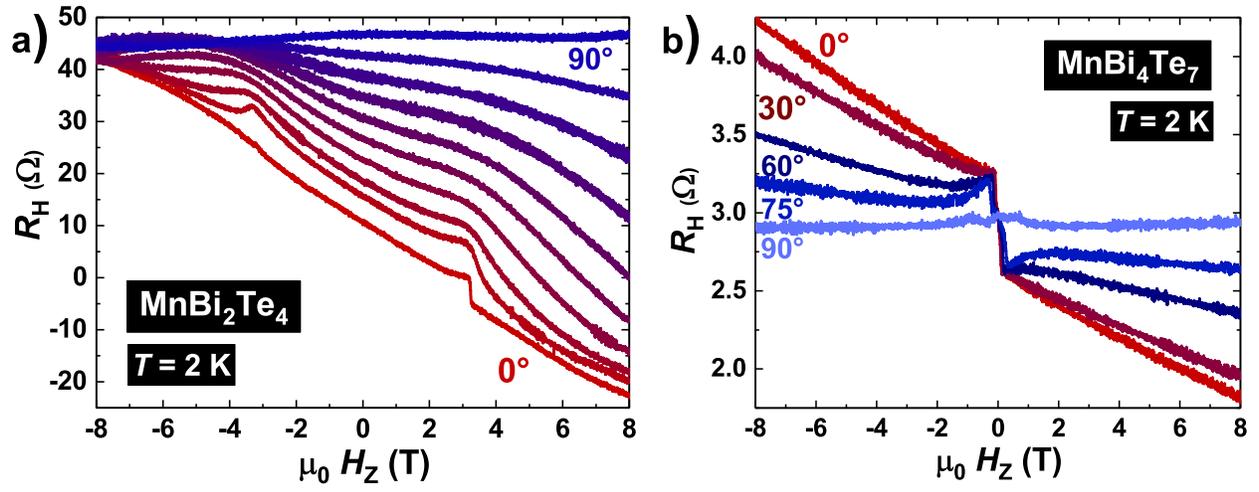}
\centering
\caption{
Angular dependence of the transverse Hall resistance for MnBi$_2$Te$_4$ (\textbf{a}), measured at $T=2$~K, and for MnBi$_4$Te$_7$ (\textbf{b}), measured at $T=8$~K.
}
\label{fig2SI}
\end{figure}

For a perpendicular field, that is, applied along the easy anisotropy axis ($\theta=0^{\circ}$), the transverse Hall resistance for MnBi$_2$Te$_4$ shows a change due to a magnetization jump at the spin-flop transition followed by the coherent rotation of the canted magnetization. This change in the $M_\textrm{Z}$ component is consistent with the evolution of the longitudinal resistance described in the main text, up to the saturation field visible as a kink at about 7~T. The latter corresponds exactly to the exchange field (for $\theta=0^{\circ}$), which is indeed much larger than the anisotropy field in MnBi$_2$Te$_4$. Besides, this magnetization reversal mechanism remains the same for all temperatures, with only a rapid change of $H_\textrm{SF}$ close to the N\'eel temperature, which is also consistent with a dominant exchange energy. As seen in Fig.~\ref{fig2SI}a) at $T=2$~K, the angular dependence of the transverse resistance has a rapid vanishing of the anomalous Hall contribution when tilting the magnetic field. As shown by our calculations (see Fig.~\ref{fig4SI}a), this is indeed typical for a small $K/J$ ratio, which confirms that the effective anisotropy constant always remains much smaller than the inter-layer exchange coupling in MnBi$_2$Te$_4$, for all temperatures.  

The situation is different for MnBi$_4$Te$_7$. As seen in Fig.~\ref{fig2SI}b) at $T=8$~K, there is a sharp magnetization jump at the spin-flop transition with a small canting and a fast evolution to the saturated magnetization state. As shown by our calculations (see Fig.~\ref{fig4SI}b), this is typical for a $K/J$ ratio close to or larger than one. When the anisotropy energy compares to the exchange energy, both sub-lattice magnetizations still rotate away from the easy-anisotropy axis at the spin-flop transition, but the anisotropy is strong enough to reduce their canting. As a consequence, the magnetization is already nearly saturated after the spin-flop event. In a tilted field, the coherent rotation becomes visible (Fig.~\ref{fig2SI}b), with an anisotropy field $\mu_0 H_\textrm{A}\approx 0.7$~T. Also, the switching field now has a progressive angle dependence, as shown in the main text, similar to the results of our calculations for a large $K/J$ ratio (see Fig.~\ref{fig4SI}b). In MnBi$_4$Te$_7$,the effective anisotropy even becomes larger than the antiferromagnetic exchange coupling at lower temperatures, so that the uniaxial anisotropy solely drives the magnetization reversal (spin-flip events).

\section{Supplementary Information: Bi-layer Stoner-Wohlfarth model}

\subsection{Longitudinal-field dependence of the free energy}

To model our experiments, we calculated the free energy of the modified Stoner-Wohlfath model discussed in the main text, for all possible configurations of the sub-lattice magnetization orientations $\theta_1$ and $\theta_2$. Both the magnetic ground state and the energy barrier to a higher-energy metastable state can be inferred from the free energy diagrams, for any amplitude and orientation of the applied magnetic field $\overrightarrow{H}$, with a polar angle $\theta$ with respect to the easy-anisotropy axis. 

\begin{figure}[!t!h]
\includegraphics[width=\columnwidth]{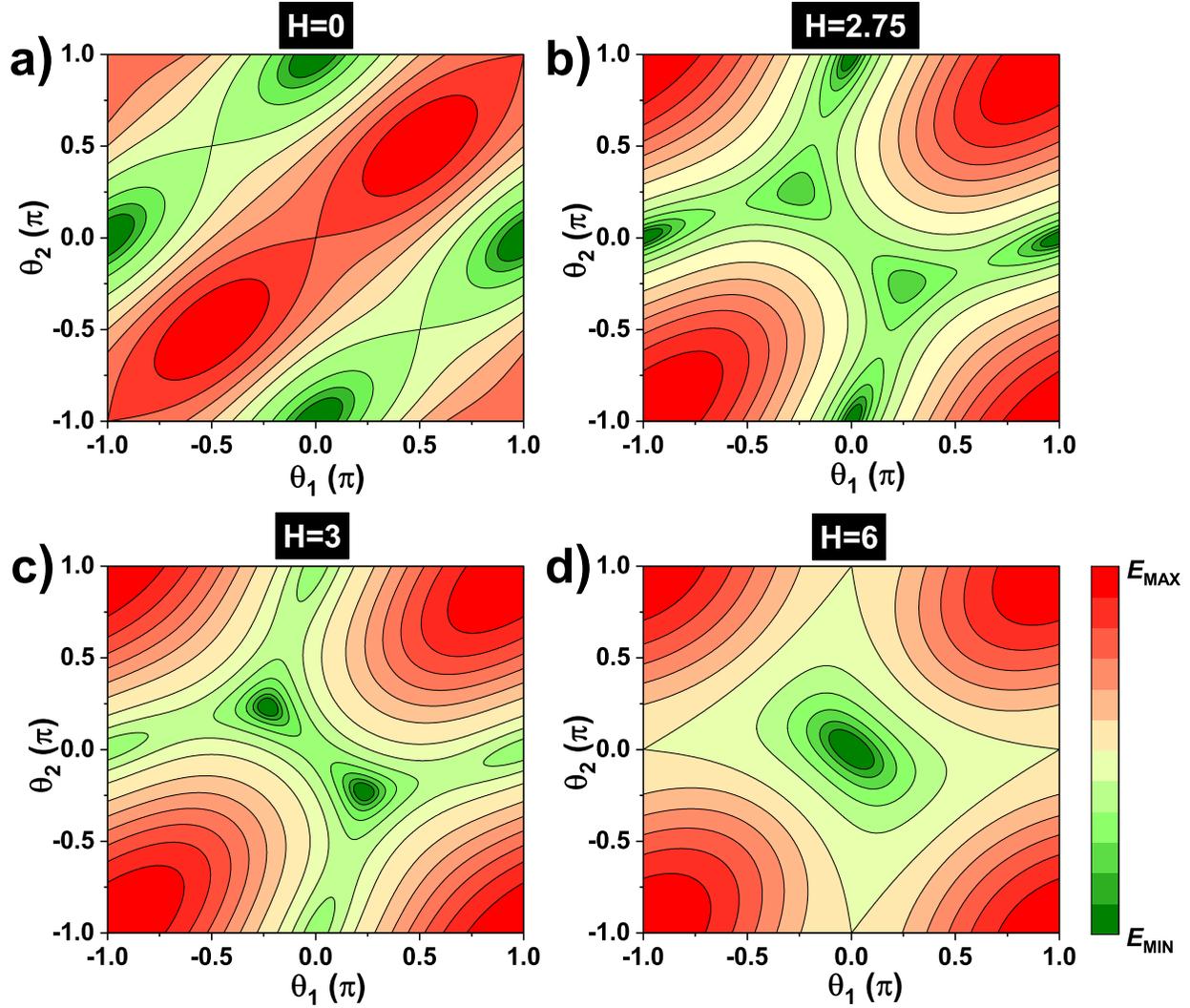}
\centering
\caption{
Free-energy contours in the sub-lattice magnetization orientation coordinates ($\theta_1$, $\theta_2$) calculated for four different values of the magnetic field $H$ applied along the anisotropy easy axis. The ground state, determined by global energy minima, changes abruptly from an anti-parallel [\textbf{a}), $H=0$; \textbf{b}), $H=2.75$] to a canted [\textbf{c}), $H=3$] magnetization configuration, which then rotates to a parallel state in large fields [\textbf{d}), $H=6$].
}
\label{fig3SI}
\end{figure}

As an example, we consider the case of a magnetic field applied along the easy-anisotropy axis, that is, perpendicularly to the sample plane ($\theta=0^{\circ}$). In zero field (Fig.~\ref{fig3SI}a), the A-type antiferromagnetic ground state corresponds to ($\theta_1$, $\theta_2$)=($0^{\circ}$,$180^{\circ}$) or ($180^{\circ}$,$0^{\circ}$). In a finite field, two metastable states develop, as shown in Fig.~\ref{fig3SI}b,c), and become the new equilibrium state at the spin-flop field. The exact values of $\theta_1$ and $\theta_2$ just after the spin-flop transition depend on the $K/J$ ratio. In a large applied field (Fig.~\ref{fig3SI}d), the sub-lattice magnetizations align parallel to each other, with ($\theta_1$, $\theta_2$)=($0^{\circ}$,$0^{\circ}$) or ($180^{\circ}$,$180^{\circ}$).

\subsection{Angular dependence of the calculated hysteresis loops}

\begin{figure}[!t!h]
\includegraphics[width=\columnwidth]{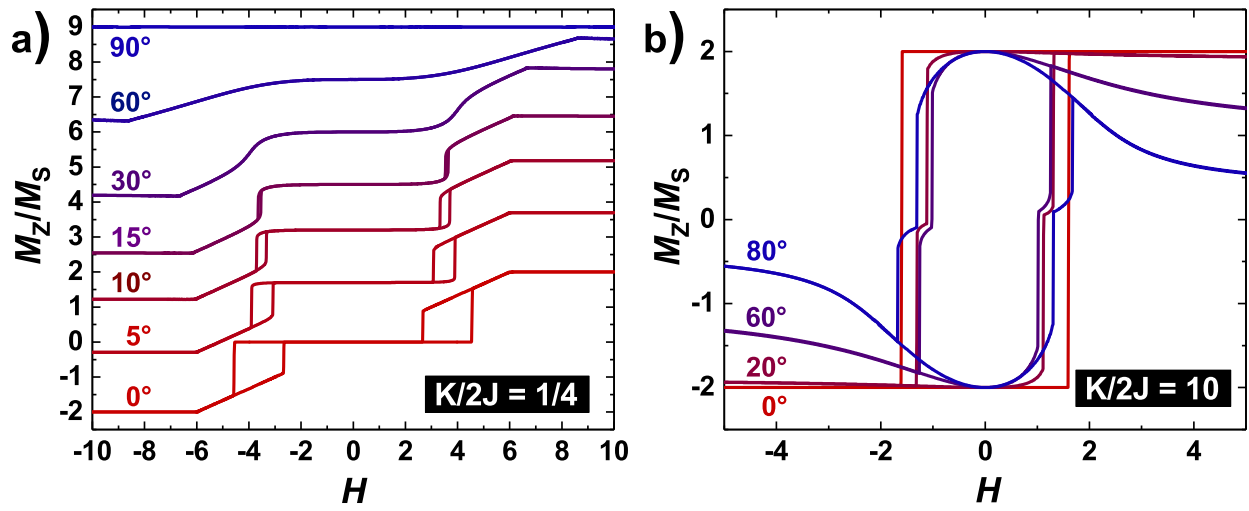}
\centering
\caption{
Angular dependence of the hysteresis loops calculated for a small $K/2J=1/4$ ratio \textbf{a}) and for a large $K/2J=10$ ratio \textbf{b}).
}
\label{fig4SI}
\end{figure}

From the field dependence of equilibrium states, we calculated the hysteresis loops for arbitrary $K/J$ ratios and magnetic field orientations $\theta$. 
We found that the angular dependence of the spin-flop field with a large canting (small $K/J$ ratio) differs from that of the spin-flop field with a small canting ($K/J\approx1$) or of the spin-flip transition (large $K/J$ ratio).
When the exchange energy dominates, it is seen that the spin-flop transition disappears rather quickly under a tilt of the applied field (Fig.~\ref{fig4SI}a), whereas it has a progressive evolution when the uniaxial anisotropy becomes comparable to or larger than the exchange energy (Fig.~\ref{fig4SI}b).  
This corresponds well to the difference found experimentally between MnBi$_.2$Te$_4$ and MnBi$_4$Te$_7$.


Furthermore, the model also captures other important aspects, such as the evolution of the magnetization under coherent rotation towards or away from the easy-anisotropy axis, as well as the angular dependence of the saturation field in the regime $K/J<1$.


\newpage

\newpage

\newpage

\newpage

\end{document}